\documentclass[twocolumn,showkeys,aps,prb,showpacs]{revtex4-1}
\usepackage{graphicx}
\usepackage[CJKbookmarks,dvipdfm,colorlinks,linkcolor=blue,citecolor=blue]{hyperref}

\begin{document}

\title{Thermoelectric properties of topological insulator $\mathrm{BaSn_2}$}

\author{San-Dong Guo and Liang Qiu}
\affiliation{Department of Physics, School of Sciences, China University of Mining and
Technology, Xuzhou 221116, Jiangsu, China}
\begin{abstract}
Recently,  $\mathrm{BaSn_2}$ is predicted to be a strong topological insulator by the first-principle calculations. It is well known that  topological insulator has a close connection  to thermoelectric  material, such as $\mathrm{Bi_2Te_3}$ family. In this work, we investigate  thermoelectric properties of $\mathrm{BaSn_2}$ by the first-principles combined with Boltzmann
transport theory. The electronic part is carried out by a modified Becke and Johnson (mBJ) exchange potential, including spin-orbit coupling (SOC), while the phonon part is performed using generalized gradient approximation (GGA).
It is found that the electronic transport coefficients between the in-plane and cross-plane directions show the strong anisotropy, while lattice lattice thermal conductivities  show an almost isotropy.  Calculated results show  a very low lattice thermal conductivity for $\mathrm{BaSn_2}$, and the corresponding average lattice thermal conductivity  at room temperature is 1.69   $\mathrm{W m^{-1} K^{-1}}$, which is comparable or lower than those of lead chalcogenides and bismuth-tellurium systems  as classic thermoelectric materials. Due to the complicated scattering mechanism, calculating scattering time $\tau$ is challenging.
By using a empirical $\tau$=$10^{-14}$ s, the n-type figure of merit $ZT$ is greater than 0.40 in wide temperature range.
Experimentally, it is possible to attain better thermoelectric performance, or to enhance one by strain or tuning  size parameter. This work indicates that  $\mathrm{BaSn_2}$  may be a potential thermoelectric  material, which can stimulate further theoretical and experimental works.
\end{abstract}


\maketitle

\section{Introduction}
Searching high-performance thermoelectric  materials is challenging and urgent, which can make  essential contributions
to energy crisis and global warming by directly converting  heat to electricity. The dimensionless thermoelectric figure of merit $ZT$, which  determines the  performance  of thermoelectric material\cite{s1000,s2000}, can be written as
\begin{equation}\label{i1}
 ZT=S^2\sigma T/(\kappa_e+\kappa_L)
\end{equation}
where S, $\sigma$, T, $\kappa_e$ and $\kappa_L$ are the Seebeck coefficient, electrical conductivity, absolute working temperature, the electronic and lattice thermal conductivities, respectively.
To attain a high $ZT$,  a high power factor ($S^2\sigma$) and a low thermal conductivity ($\kappa=\kappa_e+\kappa_L$) are required.  Unfortunately, Seebeck coefficient and   electrical conductivity are generally coupled with each other, and they are oppositely  proportional to carrier density. So, an appropriate carrier density, neither too big nor too small, can lead to the maximum power factor, and then narrow band-gap materials are potential for efficient thermoelectric applications\cite{s3}.
Topological insulators, characterized by a full insulating bulk gap and gapless edge states\cite{s4,s5}, share similar material properties with thermoelectric  materials, such as heavy elements and narrow gaps. For example bismuth-tellurium systems,
they are both famous topological insulators\cite{s6} and good thermoelectric materials\cite{s7}.
It has been demonstrated that $ZT$ of topological insulators is strongly size dependent, and can be improved to be greater than 1 by tuning size parameter\cite{s8}.

\begin{figure}
  \includegraphics[width=7.0cm]{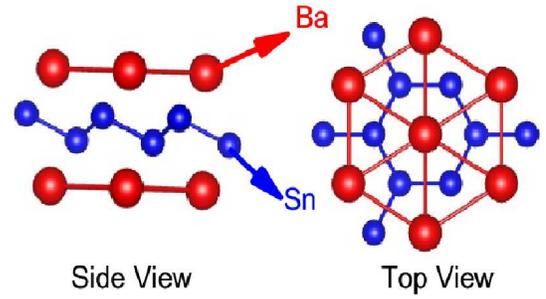}
  \caption{The crystal structures of $\mathrm{BaSn_2}$, which is composed of alternating  buckled honeycomb Sn layers  and
flat triangular Ba layers.}\label{struc}
\end{figure}
\begin{figure}[htp]
  \includegraphics[width=8cm]{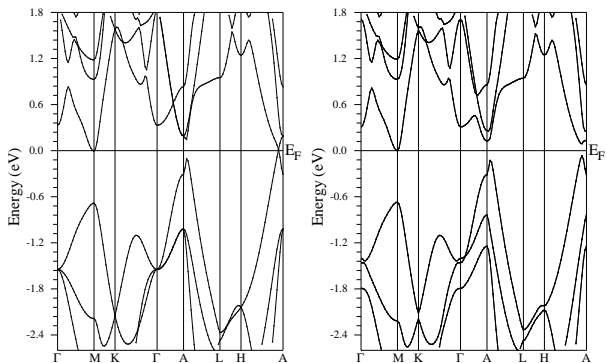}
  \caption{The energy band structures of $\mathrm{BaSn_2}$  using mBJ (Left) and mBJ+SOC (Right).}\label{band}
\end{figure}

Recently, $\mathrm{BaSn_2}$ has been predicted to be  a new strong topological insulator\cite{s9,s10}, which is composed of alternating  buckled honeycomb Sn layers  and flat triangular Ba layers. A topological band inversion is  at high symmetry A point, which is different from known strong topological insulators inverting  at high symmetry $\Gamma$ point. It is interesting and urgent to investigate thermoelectric properties of $\mathrm{BaSn_2}$. Another important cause is that $\mathrm{BaSn_2}$  contains buckled honeycomb Sn layers known as stanene,  which has been
realized via molecular beam epitaxy on a $\mathrm{Bi_2Te_3}$ substrate\cite{s11}.
Stanene  with buckled honeycomb structure like silicene\cite{s111} and germanene\cite{s112} has been also predicted
 to be  a topological insulator,  supporting  a large-gap two-dimensional (2D) quantum spin Hall state\cite{s12}.
The thermoelectric properties of Stanene also have been investigated\cite{s8,s13}, and the corresponding room-temperature lattice thermal conductivity  is 11.6 $\mathrm{W m^{-1} K^{-1}}$\cite{s13}.  It is also interesting to know whether Ba layers can produce reduced effects on lattice thermal conductivity.

Here, we investigate  the thermoelectric properties of  $\mathrm{BaSn_2}$, including both electron and phonon parts.
Calculated results show that the electronic transport coefficients along a and c axises  show the strong anisotropy.
At room temperature, the lattice thermal conductivity is 1.77 $\mathrm{W m^{-1} K^{-1}}$ along a axis and 1.54 $\mathrm{W m^{-1} K^{-1}}$ along c axis, respectively, which is lower than one (11.6 $\mathrm{W m^{-1} K^{-1}}$) of Stanene, and which is also comparable or lower than those of good thermoelectric materials, such as lead chalcogenides and bismuth-tellurium systems.
The additional result is that the structural stability of $\mathrm{BaSn_2}$ is proved by phonon
dispersion and mechanical stability criterion within elastic constants.
According to calculated $ZT$ values, it is found that n-type doping may provide better thermoelectric performance.

\begin{figure*}[htp]
  \includegraphics[width=12cm]{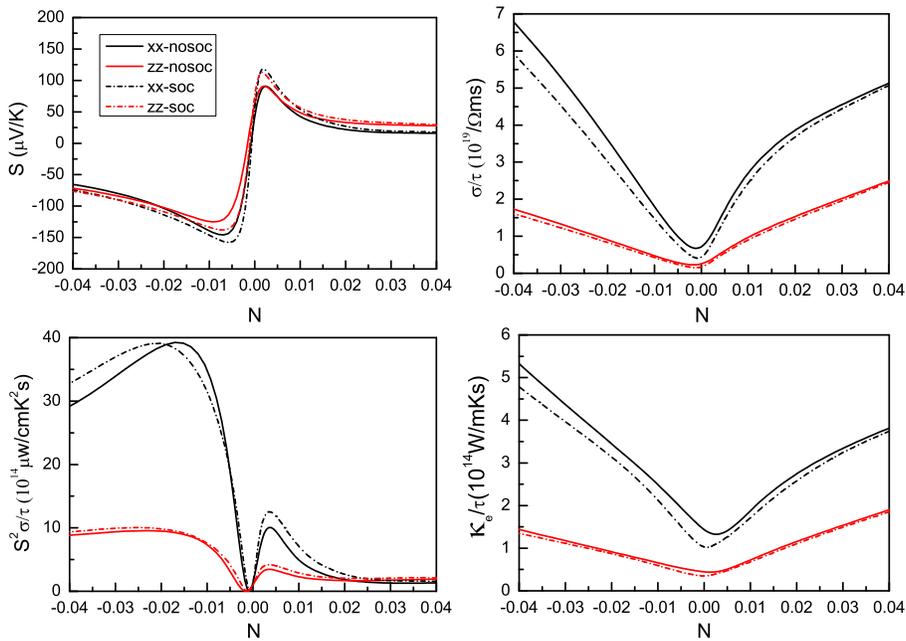}
  \caption{(Color online) At room temperature (300 K),  transport coefficients  as a function of doping level (N) along a and c axises:  Seebeck coefficient S,  electrical conductivity with respect to scattering time  $\mathrm{\sigma/\tau}$,  power factor with respect to scattering time $\mathrm{S^2\sigma/\tau}$  and electronic thermal conductivity with respect to scattering time $\mathrm{\kappa_e/\tau}$ calculated with mBJ  and mBJ+SOC. }\label{s0}
\end{figure*}

\begin{figure}[htp]
  \includegraphics[width=7cm]{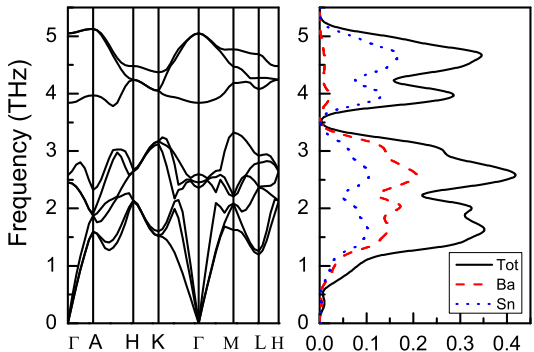}
  \caption{Phonon band structure with phonon DOS of $\mathrm{BaSn_2}$ using GGA-PBE.}\label{ph}
\end{figure}

\begin{figure}[htp]
  \includegraphics[width=7cm]{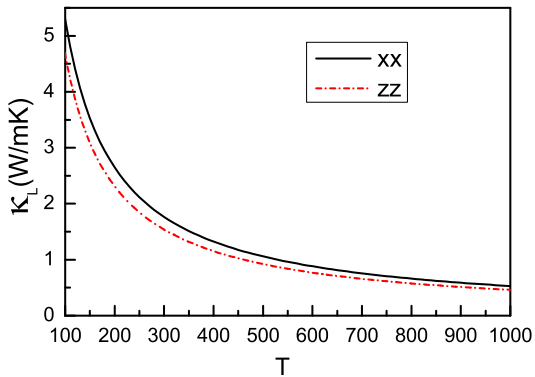}
  \caption{The lattice thermal conductivities of $\mathrm{BaSn_2}$ along  a and c axises  using GGA-PBE.}\label{kl}
\end{figure}

\begin{figure}[htp]
  \includegraphics[width=8cm]{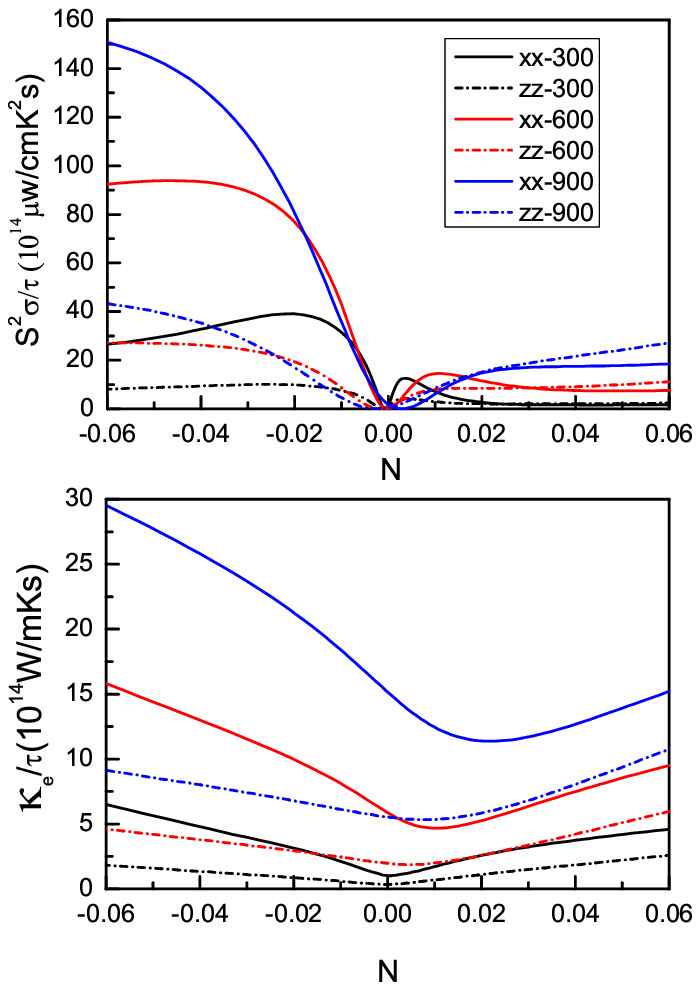}
  \caption{The power factor with respect to scattering time $\mathrm{S^2\sigma/\tau}$  and electronic thermal conductivity with respect to scattering time $\mathrm{\kappa_e/\tau}$ along  a and c axises as a function of doping level with temperature  being 300, 600 and 900 K.}\label{s1}
\end{figure}

The rest of the paper is organized as follows. In the next section, we shall
describe computational details for first-principle and transport coefficients calculations. In the third section, we shall present the electronic structures and  thermoelectric properties of $\mathrm{BaSn_2}$. Finally, we shall give our discussions and conclusion in the fourth section.

\section{Computational detail}
 A full-potential linearized augmented-plane-waves method
within the density functional theory (DFT) \cite{1} is employed to study electronic structures of $\mathrm{BaSn_2}$, as implemented in the package WIEN2k \cite{2}. An improved  Tran and Blaha's mBJ
 exchange potential plus local-density approximation (LDA) correlation potential for the
exchange-correlation potential \cite{4} is employed  to  produce
more accurate band gaps. The  free  atomic position parameters  are optimized using GGA of Perdew, Burke and  Ernzerhof  (GGA-PBE)\cite{pbe} with a force standard of 2 mRy/a.u..
 The full relativistic effects are calculated
with the Dirac equations for core states, and the scalar
relativistic approximation is used for valence states
\cite{10,11,12}. The SOC was included self-consistently
by solving the radial Dirac equation for the core electrons
and evaluated by the second-variation method\cite{so}. The convergence results are determined
by using  6000 k-points in the
first Brillouin zone for the self-consistent calculation, making harmonic expansion up to $\mathrm{l_{max} =10}$ in each of the atomic spheres, and setting $\mathrm{R_{mt}*k_{max} = 8}$ for the plane-wave cut-off. The self-consistent calculations are
considered to be converged when the integration of the absolute
charge-density difference between the input and output electron
density is less than $0.0001|e|$ per formula unit, where $e$ is
the electron charge. Based on the results of electronic
structure, transport coefficients for electron part
are calculated through solving Boltzmann
transport equations within the constant
scattering time approximation (CSTA),  as implemented in
BoltzTrap\cite{b}, which shows reliable results in many classic thermoelectric
materials\cite{b1,b2,b3}. To
obtain accurate transport coefficients, we set the parameter LPFAC for 10, and use 32000 k-points (35$\times$35$\times$25 k-point mesh) in the first Brillouin zone for the energy band calculation. The  lattice thermal conductivities are performed
by using Phono3py+VASP codes\cite{pv1,pv2,pv3,pv4}. For the third-order force constants, 2$\times$2$\times$2 supercells
are built, and reciprocal
spaces of the supercells are sampled by  4$\times$4$\times$3 meshes. To compute lattice thermal conductivities, the
reciprocal spaces of the primitive cells  are sampled using the 20$\times$20$\times$19 meshes.
 \begin{table}[!htb]
\centering \caption{ Lattice thermal conductivities (Unit:$\mathrm{W m^{-1} K^{-1}}$) of  $\mathrm{BaSn_2}$, lead chalcogenides\cite{lc1} and bismuth-tellurium systems\cite{lc2,lc3} at 300 K.}\label{tab0}
 \begin{tabular*}{0.48\textwidth}{@{\extracolsep{\fill}}ccccccc}
  \hline\hline

 $\mathrm{BaSn_2}$ & PbS & PbSe&  PbTe&$\mathrm{Bi_2Te_3}$ & $\mathrm{Sb_2Te_3}$& $\mathrm{Bi_2Se_3}$\\\hline\hline
 1.69 &2.9  &2.0 & 2.5& 1.6&2.4&1.34\\\hline\hline
\end{tabular*}
\end{table}

\section{MAIN CALCULATED RESULTS AND ANALYSIS}
 $\mathrm{BaSn_2}$  crystallizes in the  $\mathrm{EuGe_2}$-type hexagonal structure with  space group  being $P\bar{3}m1$ (No.164), which  is composed of alternating  buckled honeycomb Sn layers  and flat triangular Ba layers (See \autoref{struc}).
  According to stacking of Sn,  the  Sn layer is actually famous stanene\cite{s11}, a monolayer  with buckled honeycomb structure like silicene and germanene\cite{s111,s112}.
In our calculations, the experimental values (a=b=4.652 $\mathrm{\AA}$, c=5.546 $\mathrm{\AA}$)\cite{labc} are used for a, b and c,
and the free atomic position of Sn is optimized within GGA-PBE. The optimized position of  Sn (1/3, 2/3, 0.107) is in good agreement with experimental value of Sn (1/3, 2/3, 0.103). An improved mBJ exchange potential is employed  to investigate  electronic structures of $\mathrm{BaSn_2}$, which is superior to GGA and LDA for gap calculations.
$\mathrm{BaSn_2}$ is predicted to be a topological insulator\cite{s9,s10}, so SOC is included for electron  part. The energy band structures of $\mathrm{BaSn_2}$ using both mBJ and mBJ+SOC are plotted \autoref{band}. It is found that SOC has observable effects on energy bands around high symmetry A point. The mBJ results show that the valence band extrema (VBE) and conduction band extrema (CBE) along H-A line almost  coincide, while a gap   of 156 meV is produced with mBJ+SOC.  Another noteworthy thing is that $\mathrm{BaSn_2}$ has many CBE with their energies being very close, which is benefit for high Seebeck coefficient\cite{s1000}.

The  transport coefficients for electron part, including  Seebeck coefficient and electrical conductivity, are carried out  using CSTA Boltzmann theory. Seebeck coefficient is independent
of scattering time, while electrical conductivity depends  on scattering time.
Within the framework of  rigid band approach, the doping level  can be achieved   by simply shifting  Fermi level, which is reasonable in the low doping level in many thermoelectric materials\cite{tt9,tt10,tt11}. The doping level is defined as  electrons (minus value) or holes (positive value) per unit cell.  The n-type doping (negative doping levels) can be simulated by shifting  Fermi level  into conduction bands, producing  the negative Seebeck coefficient.
 When the Fermi level moves into valence bands, the p-type doping (positive doping levels) with the positive Seebeck coefficient can be achieved.

 Due to crystal symmetry of $\mathrm{BaSn_2}$, the physical properties along a and b axises (the in-plane  direction) are equivalent, which are different from those along c axis (the cross-plane direction), so we only show  transport coefficients along a and c axises. At 300 K, the doping level dependence of   Seebeck coefficient S,  electrical conductivity with respect to scattering time  $\mathrm{\sigma/\tau}$,  power factor with respect to scattering time $\mathrm{S^2\sigma/\tau}$  and   electrical thermal  conductivity with respect to scattering time  $\mathrm{\kappa_e/\tau}$ along  a and c axises  using mBJ and mBJ+SOC are plotted in \autoref{s0}.
 It is found that SOC has a enhanced effect on Seebeck coefficient, which can be explained by that some VBE or CBE are more close to each other  at the presence of SOC. But, SOC has a reduced influence on  electrical conductivity and electrical thermal  conductivity.  Calculated results show that SOC has a mildly improved effect on power factor, except for n-type one along a axis in low doping. The n-type doping  has larger Seebeck coefficient than the p-type one, which is because the numbers of CBE with adjacent energy are more than ones of VBE, and then n-type power factor is very larger than p-type one.
 The anisotropy of Seebeck coefficient along a and c axises  is tiny, while electrical conductivity shows strong anisotropy.
The  electrical conductivity along a axis is almost  three times as large as one along c axis, which  implies the in-plane  direction as a much more conductive direction.
 The power factor and  electrical thermal  conductivity  show the same anisotropy with electrical conductivity. It is observed that electrical thermal  conductivity has similar trend with  electrical conductivity, which is due to the Wiedemann-Franz law: $\mathrm{\kappa_e}$= $L$$\mathrm{\sigma}$$T$, where $L$ is the Lorenz number.

The phonon band structure with phonon density of states  (DOS) of $\mathrm{BaSn_2}$ using GGA-PBE  are shown in \autoref{ph}.
The unit cell of  $\mathrm{BaSn_2}$  contains two  Sn and one Ba atoms, resulting in 3 acoustic
and 6 optical phonon branches. The maximum optical frequency of 5.13 THz is comparable with  the corresponding
values  of good thermoelectric materials, such as  $\mathrm{Bi_2Te_3}$\cite{ph1} (4.5 THz) and SnSe\cite{ph2} (5.6 THz).
The maximal acoustic vibration frequency  in  $\mathrm{BaSn_2}$ is only 2.36 THz, which  is benefit to low thermal conductivity. It is found that
the phonon dispersions are separated into two regions  with a gap around 0.47 THz.
According to projected DOS in Ba and Sn atoms, it can be noted that the lower (upper) part of phonon dispersions mainly is due to the vibrations of the heavy Ba (light Sn) atoms. Calculated results show that  no imaginary frequency
modes are produced,  suggesting no structural instability at low temperature for $\mathrm{BaSn_2}$.

 Based on the following mechanical stability criterion for a rhombohedral structure\cite{el1}:
\begin{equation}\label{1a}
    C_{11}-|C_{12}|>0
\end{equation}
\begin{equation}\label{1a}
    (C_{11}+C_{12})C_{33}-2C_{13}^2>0
\end{equation}
\begin{equation}\label{1a}
     (C_{11}-C_{12})C_{44}-2C_{14}^2>0
\end{equation}
The $C_{ij}$ are five independent elastic constants, and the calculated values are listed in \autoref{tab1}.
It is easy to conclude that these criteria are satisfied for $\mathrm{BaSn_2}$, implying  mechanical stability
of $\mathrm{BaSn_2}$.  Based on calculated elastic constants $C_{ij}$,   the bulk modulus $B$,  shear modulus $G$, Young's modulus $E_x$ and $E_z$ can be attained, and list them in \autoref{tab1}. $B/G$ can measure  ductility and brittleness, and the critical value separating ductile and brittle materials is  $\sim$1.75\cite{el1}. The calculated ratio is 1.89, indicating the ductile nature of $\mathrm{BaSn_2}$. The anisotropy ratios also can be calculated by the following expressions:
\begin{equation}\label{1a}
    A_1=2C_{44}/(C_{11}-C_{12})
\end{equation}
\begin{equation}\label{1a}
   A_2=4C_{44}/(C_{11}+C_{33}-2C_{13})
\end{equation}
The numeric values of $A_1$ and $A_2$ are 1.39 and 1.64, which are close to unity, suggesting relatively small anisotropy. The related data  are also summarized in \autoref{tab1}.  Moreover, it is very interesting that the elastic parameters of $\mathrm{BaSn_2}$ are very close to ones of good thermoelectric material  $\mathrm{Bi_2Te_3}$\cite{el2}, which may mean they share similar lattice thermal conductivity.

\begin{table}[!htb]
\centering \caption{The calculated elastic constants $C_{ij}$ , bulk modulus $B$ , shear modulus $G$ and  Young's modulus $E_x$, $E_y$ (Their Unit: GPa); $B/G$ and anisotropy factor $A_1$, $A_2$.}\label{tab1}
  \begin{tabular*}{0.48\textwidth}{@{\extracolsep{\fill}}ccccccc}
  \hline\hline

$C_{11}$ & $C_{12}$ &$C_{13}$& $C_{14}$&$C_{33}$ & $C_{44}$& $B$\\\hline
71.89 &31.81  &23.39 & 12.46& 43.10&27.91&36.63\\\hline\hline
$G$ & $E_x$ &$E_z$&$B/G$&$A_1$&$A_2$\\\hline
19.39&42.28&32.55&1.89&1.39&1.64 \\\hline\hline
\end{tabular*}
\end{table}

\begin{table*}[!htb]
\centering \caption{ Peak $ZT$ along a and c axises for both n- and p-type  with $\tau$=$10^{-14}$ s, and the corresponding doping concentrations. The doping concentration equals  $\mathrm{9.621\times10^{21}cm^{-3}}$ $\times$ doping level. }\label{tab2}
  \begin{tabular*}{0.96\textwidth}{@{\extracolsep{\fill}}ccccccccc}
  \hline\hline

                 &    & a-axis  & & &   &  c-axis& &\\\hline
                   &n&&p&&n&&p&\\
  T (K) &($\mathrm{\times10^{19}cm^{-3}}$)&$ZT$&($\mathrm{\times10^{19}cm^{-3}}$)&$ZT$&($\mathrm{\times10^{19}cm^{-3}}$)&$ZT$&($\mathrm{\times10^{19}cm^{-3}}$)&$ZT$\\\hline\hline
300 &13.99&0.25&3.26&0.13&17.53&0.12&3.26&0.06\\
600 &25.32&0.44&10.55&0.16&31.93&0.35&10.55&0.17\\
900&45.99&0.46&28.58&0.13&57.49&0.40&29.75&0.23\\\hline\hline
\end{tabular*}
\end{table*}
\begin{figure}[htp]
  \includegraphics[width=7cm]{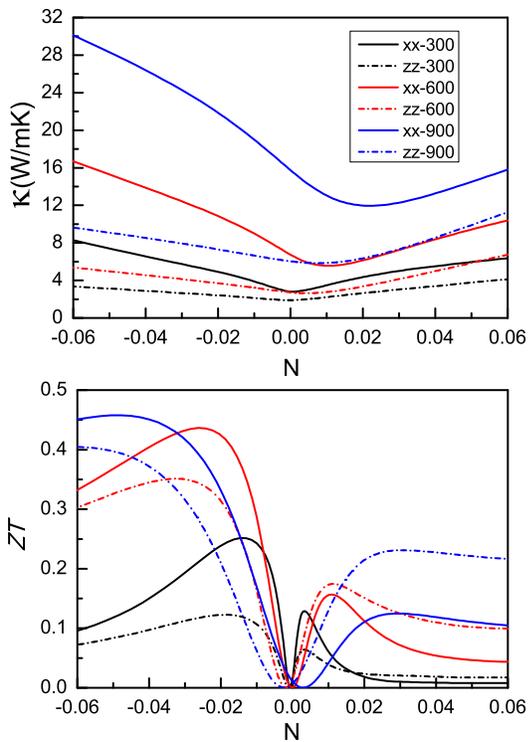}
  \caption{The thermal conductivity $\kappa$=$\kappa_e$+$\kappa_L$ and $ZT$ along  a and c axises as a function of doping level with temperature  being 300, 600 and 900 K, and the scattering time $\mathrm{\tau}$  equals 1 $\times$ $10^{-14}$ s.}\label{s2}
\end{figure}

 Based on harmonic and anharmonic
interatomic force constants, the lattice thermal conductivities of $\mathrm{BaSn_2}$ can be obtained, which  along a and c axises  as a function of temperature are plotted \autoref{kl}. It is  assumed that the lattice thermal conductivity is independent of  doping level, and typically goes as 1/T, which can be found in many thermoelectric materials\cite{lc11,lc21}.
It is found that the lattice thermal conductivity
 exhibits  little anisotropy, where the lattice thermal conductivity
along c axis is lower than that along  a axis.  The  room-temperature lattice thermal conductivity is 1.77 $\mathrm{W m^{-1} K^{-1}}$ along a axis and 1.54 $\mathrm{W m^{-1} K^{-1}}$ along c axis, respectively, which is very lower than that of (11.6 $\mathrm{W m^{-1} K^{-1}}$) of Stanene\cite{s13} due to Ba layers.
The average lattice thermal conductivity 1.69 $\mathrm{W m^{-1} K^{-1}}$ along three axises can match with ones of lead chalcogenides and bismuth-tellurium systems  as classic thermoelectric materials\cite{lc1,lc2,lc3}, and we summary related lattice thermal conductivities in \autoref{tab0}. The lattice thermal conductivity of  $\mathrm{BaSn_2}$ is even lower than ones of conventional good thermoelectric
materials except for  $\mathrm{Bi_2Se_3}$ and  $\mathrm{Bi_2Te_3}$.
 Such a low lattice thermal conductivity implies $\mathrm{BaSn_2}$ may be a potential thermoelectric material.

Finally, we consider  possible efficiency of thermoelectric conversion based on calculated  transport coefficients of electron and phonon parts. The power factor and  electronic thermal conductivity with respect to scattering time $\mathrm{S^2\sigma/\tau}$   and $\mathrm{\kappa_e/\tau}$ along  a and c axises as a function of doping level with temperature  being 300, 600 and 900 K are plotted in \autoref{s1}. To attain the figure of merit $ZT$, a unknown parameter is scattering time $\mathrm{\tau}$.  Calculating  scattering time $\mathrm{\tau}$  from the first-principle calculations is difficulty  because of the complexity of various carrier scattering mechanisms. If some experimental transport coefficients are available, the scattering time  can be attained by comparing the experimental transport coefficients with calculated ones, such as electrical conductivity. Unfortunately, the related  experimental transport coefficients of $\mathrm{BaSn_2}$ are unavailable, and we use a empirical scattering time 1 $\times$ $10^{-14}$ s to estimate possible $ZT$ values. The thermal conductivity $\kappa$=$\kappa_e$+$\kappa_L$ and $ZT$ along  a and c axises as a function of doping level with temperature  being 300, 600 and 900 K are shown in \autoref{s2}.
In n-type doping, the  figure of merit  along a axis is always larger than the  figure of merit  along c axis.  In p-type doping,
the  a-axis figure of merit is  larger than the c-axis figure of merit at low temperature, and c-axis figure of merit becomes more larger with the increasing temperature. It is found that n-type doping exhibits more superior
thermoelectric performance than p-type doping due to more higher n-type power factor.
The peak $ZT$ along a and c axises  and  corresponding doping concentrations at three different temperatures for both n- and p-type  are shown in \autoref{tab2}. It is found that doping concentration of peak $ZT$ increases with increasing temperature.
In n-type doping, at 900 K, the a-axis $ZT$ is as high as 0.46, and 0.40 for c-axis  $ZT$. These results make us believe that $\mathrm{BaSn_2}$ may be a potential thermoelectric material.

\section{Discussions and Conclusion}
As is well known, energy band gap produces important effects on transport coefficients for electron part. Narrow gap with high carrier mobility favours  the high electrical conductivity\cite{d1}, but gapless behavior leads to a vanishing Seebeck coefficient\cite{d2}.
Seebeck coefficient and electrical conductivity are oppositely proportional to carrier density.
The electronic thermal conductivity is proportional to electrical conductivity by  Wiedemann-Franz law.
An upper limit of $ZT$ can be expressed as $ZT_e=S^2\sigma T/\kappa_e$, neglecting lattice thermal conductivity.
So, an appropriate energy band gap is needed to attain high $ZT_e$, like $\mathrm{Bi_2Te_3}$ (0.15 eV)\cite{d3}, but the mBJ+SOC gap of  $\mathrm{BaSn_2}$ is only 0.065 eV. At 300 K, the best n-type  Seebeck coefficient (about 150 $\mu$V/K) is very weaker than p-type one (about 300  $\mu$V/K) and n-type one (about 250  $\mu$V/K) of $\mathrm{Bi_2Te_3}$\cite{d3}, due to too small energy band gap.
However, narrow energy band  gap induces  high electrical conductivity for $\mathrm{BaSn_2}$, which also indicates high electronic thermal conductivity. Due to high electrical conductivity, the average best n-type power factor (about 25$\times$$10^{14}$$\mu$$\mathrm{W/cmK^2s}$) of $\mathrm{BaSn_2}$ is larger than best p-type one  (about 15$\times$$10^{14}$$\mu$$\mathrm{W/cmK^2s}$) of $\mathrm{Bi_2Te_3}$ at room temeperature\cite{d3}.  In spite of higher power factor, the average best n-type $ZT_e$ (about 0.42) is lower than best p-type one (about 0.70) of $\mathrm{Bi_2Te_3}$ at 300 K\cite{d4}, because of higher electronic thermal conductivity.

For gap calculations, mBJ is usually superior to GGA or LDA, but it may still underestimate the energy band gap of $\mathrm{BaSn_2}$. So, $\mathrm{BaSn_2}$ may have better thermoelectric performance. The energy band gap also can be tuned by strain. For example
LaPtBi\cite{d4}, when a stretched uniaxial strain is applied, it changes from semimetal into real topological insulator, achieving  comparable thermoelectric performance
with $\mathrm{Bi_2Te_3}$ by an 8\% stretched uniaxial strain. It is possible for $\mathrm{BaSn_2}$ to achieve enhanced thermoelectric performance by strain. In ref.\cite{s8}, Xu et al. point out that  $ZT$ is no longer
an intrinsic material property in topological insulator, but is strongly size dependent. Tuning size parameter can
dramatically increase $ZT$,  being significantly greater than 1.
Here, we have ignored the gapless edge states contribution to conducting channels, which helps to great improvement of thermoelectric performance by tuning size parameter.
Experimentally, it is possible to achieve further enhanced
thermoelectric performance of $\mathrm{BaSn_2}$ by tuning size parameter.

In summary,  the thermoelectric properties  of  $\mathrm{BaSn_2}$, including both electron and phonon parts, are investigated  based mainly on the reliable first-principle calculations.
It is found that SOC has little influence on electronic   transport coefficients. Calculated results show   obvious
anisotropy of  power factor and electronic thermal conductivity between the in-plane and cross-plane directions, while a slight anisotropy of lattice thermal conductivity is observed. The low lattice thermal conductivity  is  predicted, and the average one is 1.69 $\mathrm{W m^{-1} K^{-1}}$ at room temperature, which highlights possibility for $\mathrm{BaSn_2}$ as a potential  thermoelectric material. From 600 K to 900 K,  the n-type figure of merit $ZT$ is up to about 0.45 with  empirical $\tau$=$10^{-14}$ s by optimizing doping level, and it is possible to achieve improved thermoelectric performance by strain and tuning size parameter in experiment.  The present
work may be useful to  encourage further theoretical and experimental efforts to achieve high thermoelectric performance of $\mathrm{BaSn_2}$.

\begin{acknowledgments}
This work is  supported by the Fundamental Research Funds for the Central Universities (2015QNA44). We are grateful to the Advanced Analysis and Computation Center of CUMT for the award of CPU hours to accomplish this work.
\end{acknowledgments}

\end{document}